\documentclass{article}
\usepackage{spconf,amsmath,graphicx,booktabs}
\usepackage{multirow}
\usepackage{url}
\usepackage[utf8]{inputenc}
\usepackage{xcolor}
\usepackage{lipsum}
\usepackage{etoolbox}
\usepackage{amssymb}
\makeatletter

\patchcmd{\@verbatim}
  {\verbatim@font}
  {\verbatim@font\small}
  {}{}
\makeatother

\title{Unsupervised pretraining transfers well across languages}

\name{Morgane Rivière$^\dagger{}$, Armand Joulin$^\dagger{}$,  Pierre-Emmanuel Mazar\'e$^\dagger{}$,  Emmanuel Dupoux$^{\dagger{}\ddagger{}*}$}
\address{$\dagger{}$Facebook AI Research, $\ddagger{}$Ecole des Hautes Etudes en Sciences Sociales
\thanks{
 * Code and data available in \protect\url{https://github.com/facebookresearch/CPC_audio}.  This is the extended reprint of: Rivière, M., Joulin, A., Mazaré, P.E. and Dupoux, E. (2020). Unsupervised pretraining transfers well across languages. \textit{in ICASSP-2020.}}
}

\graphicspath{{figures/}}

\begin{document}

\maketitle
\begin{abstract}
  Cross-lingual and multi-lingual training of Automatic Speech Recognition (ASR) has been extensively investigated in the supervised setting.
  This assumes the existence of a parallel corpus of speech and orthographic transcriptions.
  Recently, contrastive predictive coding (CPC) algorithms have been proposed to pretrain ASR systems with unlabelled data.
  In this work, we investigate whether unsupervised pretraining transfers well across languages.
  We show that a slight modification of the CPC pretraining extracts features that transfer well to other languages, being on par or even outperforming supervised pretraining.
  This shows the potential of unsupervised methods for languages with few linguistic resources.
\end{abstract}
\begin{keywords}
Unsupervised pretraining, low resources, cross-lingual
\end{keywords}
%

\section{Introduction}
\label{sec:intro}
\vspace{-5pt}

Learning phoneme representations remains a challenge for a large number of languages with limited supervised resources.
A common approach is to pre-train these representations on a large supervised corpus in other languages and transfer them to the low resource languages~\cite{heigold2013multilingual,huang2013cross}.
For example, Vesely et al.~\cite{vesely2012language} learn a shared representation on a supervised multilingual dataset and finetune it on the target language.
This pre-training works even between distant languages, but requires massive supervised corpora in the same domain.

Recently, several works~\cite{DBLP:journals/corr/abs-1807-03748,hyvarinen2016unsupervised} have proposed promising methods to train monolingual audio representations without supervision.
In particular, Schneider et al.~\cite{schneider2019wav2vec} shows that the unsupervised pre-training method of van den Oord~\cite{DBLP:journals/corr/abs-1807-03748} improves the quality of automatic speech recognition (ASR) on several competitive benchmarks.
In this paper, we are interested to see if similar unsupervised pre-training methods can be leveraged in a cross-lingual setting to improve the quality of phoneme representations for low resource languages.

We focus on the contrastive predictive coding (CPC) method of van den Oord~\cite{DBLP:journals/corr/abs-1807-03748} since Schneider et al.~\cite{schneider2019wav2vec} has shown its benefit for pre-training features for ASR.
CPC is a form of forward modeling in the feature space~\cite{chopra2005learning}: it predicts the near future windows in an audio sequence while contrasting with windows from other sequences or more distant in time.
We introduce several modifications to the original approach stabilize the training and lead to better phoneme representations.
We use our modified CPC model to pre-train phoneme representations in English, namely on Librispeech, and transfer them to several low-resource languages from the Common Voice database.

In this paper, we obtain several results related to transferring across languages the features pre-trained without supervision.
First, pre-training phoneme representation outperforms representations trained from scratch in the target language, even if we do not use any supervision for the pre-training.
Suprisingly, we also observe that the gap between unsupervised and supervised pre-training is relatively small if we use the same pre-training corpora.
Finally, scaling unsupervised pre-training to larger unlabelled datasets further reduces the gap with the supervised pre-training features, and even surpasses it in some low-resource languages.

\vspace{-5pt}
\section{Related work}
\label{sec:related}
\vspace{-5pt}

\subsection{Multilingual pre-training for speech recognition}
A common way to improving speech recognition in low-resource languages is to train multilingual speech recognition with shared components~\cite{schultz2001language, stolcke2006cross,huang2013cross}. 
For example, Stolcke et al.~\cite{stolcke2006cross} train features for phoneme classification in a different language. 
Burget et al.~\cite{burget2010multilingual} shares the parameters of a Gaussian Mixture Model.
Closer to our work, several works have shared the parameters of a neural network encoder, using feedforward networks~\cite{vesely2012language,heigold2013multilingual,huang2013cross} or LSTM~\cite{li2019multilingual}.
The model is then finetuned on the target low-resource language to fit its specificities~\cite{dalmia2018sequence}.
The sampling of the languages during the pre-training can focus on languages related to the targeted language~\cite{li2019multilingual}.
Another approach is to encourage a language-independent encoder with an adversarial loss~\cite{adams2019massively}.
As opposed to our work, this line of research focuses on supervised pre-training which restrict its impact to domains or languages with large resources for supervision.

\vspace{-5pt}

\subsection{Unsupervised learning of features}
Many unsupervised learning approaches have been proposed for speech and we focus on those based on contrastive learning~\cite{chopra2005learning,weinberger2009distance,schroff2015facenet}.
In particular, Time Contrastive Learning~\cite{hyvarinen2016unsupervised} learns audio features by discriminating between time windows.
Our work closely follows van den Oord et al.~\cite{DBLP:journals/corr/abs-1807-03748} where a contrastive loss is used to predict forward representations in an audio sequence.
Their Contrastive Predictive Coding (CPC) objective function is similar to the objective of word2vec~\cite{mikolov2013distributed}, applied to sequences instead of words.
Contrastive approaches are also related to examplar self-supervision~\cite{dosovitskiy2014discriminative,bojanowski2017unsupervised,hjelm2018learning}. 
However, CPC has the advantage of making no assumption about the nature or number of the training data samples.
Recently, variants of CPC have been applied to monolingual ASR~\cite{schneider2019wav2vec} and images~\cite{tian2019contrastive}.

\vspace{-5pt}
\section{Approach}
\vspace{-5pt}

In this section, we rapidly introduce the approach of van den Oord et al.~\cite{DBLP:journals/corr/abs-1807-03748} and we refer the reader to the original paper for details.
We also present several modifications to improve the resulting representations and stabilize the training.
We made our code as well as our experiments available to the public\footnote{\url{https://github.com/facebookresearch/CPC_audio}}. 

\vspace{-5pt}
\subsection{Contrastive Predictive Coding}
\vspace{-5pt}

Unsupervised training of neural networks relies on building a pretext task that requires discriminative features to be solved.
The pretext task used in Contrastive Predictive Coding (CPC)~\cite{DBLP:journals/corr/abs-1807-03748} is forward modeling, i.e., predicting the future states of a sequence from its past.
The particularity of CPC is to frame forward modeling as a reconstruction of future representations, not future inputs.
Past and future representations are built from the same model, and a contrastive loss ensures that temporally nearby representations are pushed closer than temporally distant ones.

More precisely, given an audio sequence splitted in $T$ discrete time steps, or windows, we embed the input signal $\mathbf{x}_t$ at each time step $t$ with a encoder.
Then, we form the current phoneme representation $\mathbf{z}_t$ by applying a sequence model to the resulting sequence of $t$ embeddings, i.e.,
$$\mathbf{z}_t  = \psi_\rho ( \phi_\theta(\mathbf{x}_1), \dots, \phi_\theta (\mathbf{x}_t)), $$
where $\phi_\theta$ is the encoder and $\psi_\rho$ is the sequence model, parametrized by $\theta$ and $\rho$ respectively.
In~CPC, the encoder is a~$5$-layer convolutional network (kernel sizes: 10,8,4,4,4, stride sizes: 5,4,2,2,2) and the sequence model is a~$1$-layer Gated Recurrent Units~(GRU).
The encoder also has a downsampling factor of~$160$, meaning that for a~$16$kHz input, each feature encodes $10$ms of audio.

Given this phoneme embedding $\mathbf{z}_t$, the pretext task in CPC is to predict the next $K$ future representations, i.e., $\phi_\theta(\mathbf{x}_{t+k})$ for $k\in\{1,\dots,K\}$.
CPC also pushes away representations from a random subset~$\mathcal{N}_t$ of negative examples, or ``distant'' windows.
Overall, the loss function at a time step $t$ is thus:
\begin{equation}
\label{eq:cpc}
   \mathcal{L}_t = - \frac{1}{K} \sum_{k=1}^K \log \left[ \frac{ \exp\left(\phi_\theta(\mathbf{x}_{t+k})^\top\mathbf{A}_{k}\mathbf{z}_t\right) }{  \sum_{\mathbf{n}\in\mathcal{N}_t} \exp\left(\phi_\theta(\mathbf{n})^\top\mathbf{A}_{k}\mathbf{z}_t\right)}\right].
\end{equation}
where $\mathbf{A}_k$ is a linear classifier.
There are many ways to sample the ``distant'' windows and we follow van den Oord et al.~\cite{DBLP:journals/corr/abs-1807-03748} by sampling negative \emph{within speaker}.
The parameters $\theta$, $\rho$ and $\mathbf{A}_{1,\dots,K}$ are learned with stochastic gradient descent.

\vspace{-5pt}
\subsection{Modifications to Contrastive Predictive Coding}
\vspace{-5pt}
\subsubsection{Stabilization of the training}
\vspace{-5pt}

We observe empirically that the training of CPC is unstable, and can converge to poor solutions.
The reason is the presence of batch normalization ~\cite{ioffe2015batch} between the layers of the encoder.
Indeed, batch normalization parameters are learned by computing statistics over the whole batch.
Since the encoder is shared across a sequence, these parameters leak information between past and future windows.
This makes minimizing eq.~(\ref{eq:cpc}) trivial when the batch normalization is activated, resulting in instability.
We fix this issue by replacing batch normalization with a channel-wise normalization that plays a similar role of conditioning internal representations.
As oppposed to batch normalization, the parameters are not shared across the sequence and do not leak information (see Supplementary Section \ref{sec:norm} 
for details).

\vspace{-5pt}
\subsubsection{Improving the model}
\vspace{-5pt}

The prediction of future representations is made by linear classifiers on top of a phoneme embedding, as shown in eq.~(\ref{eq:cpc}).
The motivation is to encourage the phoneme embeddings to encode linearly separable phonemes.
However, the future representations are not phoneme representations themselves; they are embeddings of the time window.
Comparing the outputs of a sequence model and an encoder with a linear classifier may not result in linearly separable phoneme representations.
Several alternatives are possible, such as adding a sequence model on the future representations.
In practice, we find that replacing each linear classifier with a $1$-layer Transformer network~\cite{vaswani2017attention} works well (see Supplementary Section \ref{sec:pred} 
for details).
This layer accesses the entire sequence of  $z_{1}, .. z_{t}$ to predict a particular $\phi(x_{t+k})$.  We also observe that reducing the dimension of convolutionnal layers from $512$ to $256$ does not impact performance while reducing memory footprint.
Finally, using an LSTM instead of a GRU slightly improves the performance.

\vspace{-5pt}
\subsection{Transferring features to phoneme classification}
\vspace{-5pt}

In this work, we evaluate the quality of phoneme representations trained with no supervision when transferred across languages.
Standard cross-lingual approaches finetune their pre-trained network on the targeted language.
While this improves the quality of the resulting representations, it does not assess the quality of the pre-trained representations.
Instead, we freeze the model after the pre-training and simply learn a linear classifier for the targeted language.
Specifically, we perform the linear classification of a concatenation of $8$ windows to match the average size of a phoneme.
We then use the CTC loss between our model predictions and the non-aligned phoneme transcriptions~\cite{ICML:CTC_loss}.
This procedure explicitly measures the linear separability of the phoneme representation, once transferred to a target language.





\vspace{-5pt}
\section{Experimental setting}
\vspace{-5pt}

\label{sec:dataset}
\subsubsection{Pre-training on the Librispeech dataset}

We pre-train models on the English Librispeech dataset (LS).
We consider both the $100$h and $360$h splits of clean data.
For the supervised pre-training model, we use the aligned phone labels provided by \cite{DBLP:journals/corr/abs-1807-03748} for Librispeech-$100$h.

\vspace{-5pt}
\subsubsection{Transferring to the Common Voice database}
\vspace{-5pt}

After the pre-training, we freeze the parameters of our models and transfer the features across languages.
We consider the common Voice database\footnote{\url{https://voice.mozilla.org}} as it comes in many languages.
We retrieve  the non-aligned phoneme transcription of each audio sample by running the open-source tool phonemizer\footnote{\url{https://gitlab.coml.lscp.ens.fr/mbernard/phonemizer}} on their corresponding text scripts.
We split our dataset between train, validation and test sets along speakers to reduce the influence of speakers on the performance of phoneme predictions.
We consider two train sets of either $1$ or $5$ hours. We will open source our train-test splits along with our code.
\vspace{-5pt}
\subsubsection{Measuring phoneme separability on Zerospeech2017}
\vspace{-5pt}

Zerospeech2017 is a dataset made to measure phoneme separability of unsupervised models in different languages.
We consider the English, Mandarin and French benchmarks and we report the ABX score on them~\cite{Schatz2013ABX}.
The ABX score measures the discriminability between phonemes by estimating the probability speech segments to be closer to one another if they encode the same phoneme than if they don't (the distance being DTW-realigned average frame-wise cosine).

\vspace{-5pt}
\section{Results}
\label{sec:results}
\vspace{-5pt}

\vspace{-5pt}
\subsection{Within-language results}
\vspace{-5pt}



In this set of experiments, we compare the original CPC with our modified version on two \emph{within-language} tasks: phoneme discriminability on the English Zerospeech2017 dataset, and phoneme linear separability on Librispeech $100$h~\cite{DBLP:journals/corr/abs-1807-03748}. In Table~\ref{tab:ABX_english}, we compare our ABX score with that of the toplines from the Zerospeech leaderboards. It is interesting to note that CPC does not perform well on this metric but our modified version is on par with the state of the art. Overall, our modified CPC surpasses the original model on phoneme classification and even matches unsupervised approaches dedicated to phoneme separability. In Table~\ref{tab:Phone_separability}, we show that our modifications to CPC leads to an improvement of 3.4 points in phoneme classification compared to the original CPC implementation. 

\begin{table}[h!]
    \centering
    \begin{tabular}{lcc}
    \toprule
         & Across & Within  \\
    \midrule
    \multicolumn{3}{l}{\textit{Trained on ZeroSpeech2017 (45 h)}} \\
    Supervised topline~\cite{dunbar2017zerospeech} &6.9 & 5.3\\
      Heck et al.~\cite{heck2017feature} & 8.7 & 6.2 \\
      Chorowski et al.~\cite{chorowski2019unsupervised} & 8.0 & 5.5\\
         \midrule
    \multicolumn{3}{l}{\textit{Trained on Librispeech-360}} \\
    CPC~\cite{DBLP:journals/corr/abs-1807-03748} &13.0 & 9.6 \\
    Modified CPC & 8.5& 6.5\\
    \bottomrule
    \end{tabular}
    \caption{\textbf{Phoneme discriminability within languages.} Within- and across-speakers ABX scores for the English Zerospeech2017 test set. We compare CPC and modified CPC trained on Librispeech-360 to the best performing models.}
    \label{tab:ABX_english}
\end{table}

\begin{table}[h!]
    \centering
    \begin{tabular}{lc}
        \toprule
         &  Phone accuracy \\
         \midrule
        Supervised topline & 76.3 \\
        \midrule
        CPC~\cite{DBLP:journals/corr/abs-1807-03748} & 65.5\\
        Modified CPC & 68.9 \\
    \bottomrule
    \end{tabular}
    \caption{
      \textbf{Phone classification within language.} Accuracy on the English LibriSpeech-$100$h dataset for a linear classifier trained on top of frozen features obtained with the original and our modified CPC model.
    }
    \label{tab:Phone_separability}
\end{table}

\begin{table*}[t]
  \centering
  \begin{tabular}{lcccccccccccccc}
  \toprule
    Model & Pretraining & Frozen&  \texttt{du} & \texttt{es} & \texttt{fr}  & \texttt{it}  & \texttt{ky} & \texttt{ru} & \texttt{sv} & \texttt{tr} & \texttt{tt} & \texttt{zh}  && Avg \\
  \midrule
  From scratch    & - & No &  84.7 & 95.9 & 95.1 & 95.0 & 81.5 & 97.7 & 86.1   & 83.1& 72.9 & 84.3 && 87.6\\
    Bottleneck~\cite{fer2017multilingually} & Babel-1070h & Yes & 47.9 & 36.6 & 48.3 & 39.0 & 38.7 & 45.2 & 52.6 &  43.4 & 42.5 & 54.3  && 44.9 \\
  Supervised & LS-100h &  Yes & 42.4 &  36.4 & 47.0 &  40.5 & 41.0 & 43.6 &  47.0 & 48.5 & 41.5 & 56.8 &&  \bf 44.5\\
  \midrule
  CPC~\cite{DBLP:journals/corr/abs-1807-03748} & LS-100h & Yes & 51.5 & 44.2 & 54.5 & 47.0 & 44.8 & 49.0 & 54.0& 54.7& 48.9 & 60.1 && 50.9\\
  Modified CPC    & LS-100h & Yes & 44.4 & 38.7 & 49.3  & 42.1 & 40.7 & 45.2 & 48.8 & 49.7 & 44.0 & 55.5 && 45.8\\
  Modified CPC    & LS-360h & Yes & 42.5 & 38.0 & 47.1  & 40.5 & 41.2 & 43.7 & 47.5 & 47.3 & 42.0 & 55.0 &&  \bf 44.5 \\
  \bottomrule
  \end{tabular}
  \caption{
    {\bf Transfer of pre-trained phoneme features across languages.}
    We pre-train the features on $100$h and $360$h of Librispeech with supervision (``Supervised'') or not (``CPC'' and ``Modified CPC'').
    We also include multilingual bottleneck features (``Bottleneck'') pre-trained on $1070$h from the Babel dataset.
    We train a linear classifier on the frozen features using $1$h of speech from the Common Voice database in different languages.
    We also report a supervised model trained entirely from scratch on the $1$h of speech.
    We report Phone Error Rate.
    The languages are: 
    Dutch (\texttt{du}),
    Spanish (\texttt{es}),
    French (\texttt{fr}),
    Italian (\texttt{it}),
    Kyrgyz (\texttt{ky}),
    Russian (\texttt{ru}),
    Sweedish (\texttt{sv}),
    Turkish (\texttt{tr}),
    Tatar (\texttt{tt}) and
    Mandarin (\texttt{zh}).
  }
  \label{tab:per_1h}
\end{table*}

\begin{table}[h]
    \centering
    \begin{tabular}{l cc c cc}
    \toprule
      & \multicolumn{2}{c}{French} && \multicolumn{2}{c}{Mandarin} \\
      \cmidrule{2-3}\cmidrule{5-6}
      & A. & W.  && A. & W.  \\
      \midrule
      \multicolumn{2}{l}{\emph{Trained within language}}\\
      Supervised topline       & 9.1  & 6.8   && 5.7 & 4.2\\
      Heck et al.~\cite{heck2017feature} &11.7  & 8.7   && 7.4 & 7.9\\
      Chorowski et al.~\cite{chorowski2019unsupervised} & 10.8 & 7.5   && 11.2 & 10.7\\
      \midrule
      \multicolumn{4}{l}{\emph{Trained on English (Librispeech-360)}}\\
    CPC~\cite{DBLP:journals/corr/abs-1807-03748}     & 18.0& 12.3  && 11.5 &10.0 \\
    Modified CPC     & 14.6 & 10.0  && 9.5 & 8.9\\
    \bottomrule
    \end{tabular}
    \caption{
      \textbf{Phoneme discriminability of unsupervised features across languages.}
      Across- (``A.'') and within-speakers (``W.'') ABX scores on French and Mandarin speech for CPC features pre-trained in English. For comparison: the best systems plus supervised topline of the Zerospeech leaderboard trained within-language.}
    \label{tab:ABX_transfer}
    \vspace{-5pt}
\end{table}

\vspace{-5pt}
\subsection{Cross-lingual transfer of phoneme features}
\vspace{-5pt}

In a first experiment, we consider the problem of phoneme classification across languages on the Common Voice database.
In Table~\ref{tab:per_1h}, we report the phone error rate (PER) for the linear classifiers trained on top of the phoneme features pretrained with and without supervision.
We also compare with a model trained from scratch on the target dataset.
The training set of each target dataset is only $1$ hour long. The model trained from scratch thus performs poorly.
On the other hand, pre-trained features significantly improve the performance in all languages, even without any finetuning.
First, on $100$ hours of librispeech, our modified CPC outperforms the original CPC by 5.4 points on average.
However, supervised pre-training still performs slightly better ($1.3$ points) than our unsupervised pre-training on the same corpus.
An advantage of unsupervised pre-training is that we can apply it to any larger unannotated dataset.
We show the benefits of this by pre-training our modified CPC on $360$ hours of unlabelled data from Librispeech and match the performance of the supervised model.
This result not only confirms the findings of~\cite{schneider2019wav2vec} but it also shows that unsupervised pre-training can match supervised pre-training with enough data (see Supplementary Section \ref{sec:suppres} with the larger Libri-light dataset \cite{kahn2020}).

In a second experiment, we compare the quality of our pre-trained features against other unsupervised methods on the Zerospeech2017.
In Table~\ref{tab:ABX_transfer}, we compare on French and Mandarin, the ABX score of our approach trained on English Librispeech with unsupervised methods trained for these languages.
Surprisingly, our English features transfered to other languages are competitive with the top lines of the leaderboard.
This result further shows that unsupervised pre-trained features generalize well across languages.

\vspace{-5pt}
\subsubsection{Impact of finetuning phoneme features}
\vspace{-5pt}

\begin{table}[h]
  \centering
  \begin{tabular}{l c c c}
  \toprule
    Model & pretraining  & frozen & finetune \\
  \midrule
    From scratch & - & -& 38.3\\
    Supervised & LS-100 &  37.6 & \bf 29.2\\
  \midrule
    CPC~\cite{DBLP:journals/corr/abs-1807-03748}       & LS-100  & 43.5 & 33.3\\
    Mod. CPC & LS-100   & 38.8 & 31.0 \\
    Mod. CPC &  LS-360  & \bf 37.2 & 30.7\\
  \bottomrule
  \end{tabular}
  \caption{\textbf{Comparison between frozen and fine-tuned features.} PER averaged over $5$ languages (Spanish, French, Italian, Russian and Tatar). The training set for each language contains $5$ hours extracted from the Common Voice database. }
  \label{tab:per_5h}
\end{table}


We also study the impact of fine-tuning the phoneme features instead of freezing them.
We use $5$ hours of speech in $5$ target languages for this experiment.
In Table~\ref{tab:per_5h}, we compare the difference between frozen features and fine-tuning.
As for the experiments on $1$h of speech, our approach is on par with supervised pre-training when the features are frozen.
We also observe a boost around $7$ performance points for all the pre-training methods when we fine-tune the features.
Our approach is still relatively competitive with supervised pre-training, but slightly worse ($-1.5$ points) on average.

\vspace{-5pt}
\section{Conclusion}
\vspace{-5pt}
\label{sec:conc}
Pre-training in a given language, with or without supervision, can produce features usable across other languages and other domains. 
Moreover, these features can be matched with a set of phonemes even with extremely low resources datasets and unaligned labels.
They are usable with a very simple linear model and can be trained at low cost.
Finally, though supervised pre-training tends to be better than the unsupervised one, the gap between them is small and can be greatly reduced with the use of a larger amount of unlabelled data. We did not attempt to push numbers in order to achieve good phone error rates in the low resource languages, as we only tested a linear separation layer for phoneme classification. Further work needs to be done to establish how these pretrained features can be best used in the low resource setting (see \cite{kawakami2020}), and with other ASR tasks \cite{kahn2020}.

\let\oldbibliography\thebibliography
\renewcommand{\thebibliography}[1]{%
  \oldbibliography{#1}%
  \setlength{\itemsep}{0pt}%
}

\vspace{-5pt}
\bibliographystyle{bib}
\bibliography{egbib}

\vspace{-5pt}
\setcounter{section}{0}
\setcounter{table}{0}
\setcounter{figure}{0}

\renewcommand\thesection{S\arabic{section}}
\renewcommand\thetable{S\arabic{table}}
\renewcommand\thefigure{S\arabic{figure}}

\vspace{-5pt}
\label{sec:conc}
\onecolumn
\section{Supplementary methods}

We describe here ablation experiments comparing our reimplementation of the original CPC model  \cite{DBLP:journals/corr/abs-1807-03748} and improvements we made to this model.

\subsection{Changing the normalization method}\label{sec:norm}

In order to make the training more stable, we replaced the batch normalization in the original model with layer normalization. The results are illustrated in Table \ref{tab:ABX_ablation_LN}.

\begin{table}[h!]
    \centering
    \begin{tabular}{lcc}
    \toprule
         & Across & Within  \\
    \midrule
    \multicolumn{3}{l}{\textit{Trained on Librispeech-100}} \\
    CPC~\cite{DBLP:journals/corr/abs-1807-03748} &13.0 & 9.6 \\
    CPC + Layer norm (LN) & \textbf{12.0}& \textbf{8.7}\\
    \bottomrule
    \end{tabular}
    \caption{\textbf{Impact of the normalization method on the phoneme discriminability}. Within- and across-speakers ABX scores for the English Zerospeech2017 test set.}
    \label{tab:ABX_ablation_LN}
\end{table}

\subsection{Choosing the right predictor design}\label{sec:pred}

We compared several alternatives to the linear prediction model initially presented in \cite{DBLP:journals/corr/abs-1807-03748}. 
We supposed that if the prediction network is too simple, then the auto-regressive network will perform a significant part of the prediction task.
Thus we though that more complex architecture would improve the quality of our output features.
The results of our experiments are compiled in Table \ref{tab:ABX_ablation_predictor}.

\begin{table}[h!]
    \centering
    \begin{tabular}{lcc}
    \toprule
         & Across & Within  \\
    \midrule
    \multicolumn{3}{l}{\textit{Trained on Librispeech-100}} \\
    CPC + LN & 12.0& 8.7\\
    CPC + LN + Conv8 &13.4 & 9.2 \\
    CPC + LN + FFD & 11.7 & 8.56\\
    CPC + LN + transformer & 9.5 & 7.3 \\
    CPC + LN + transformer + dropout & \textbf{9.3} & \textbf{6.8}\\
    \bottomrule
    \end{tabular}
    \caption{\textbf{Phoneme discriminability for various predictors design}. Within- and across-speakers ABX scores for the English Zerospeech2017 test set.}
    \label{tab:ABX_ablation_predictor}
\end{table}

\section{Supplementary results}\label{sec:suppres}

\begin{figure}[b]
  \centering
  \centerline{\includegraphics[width=8.0cm]{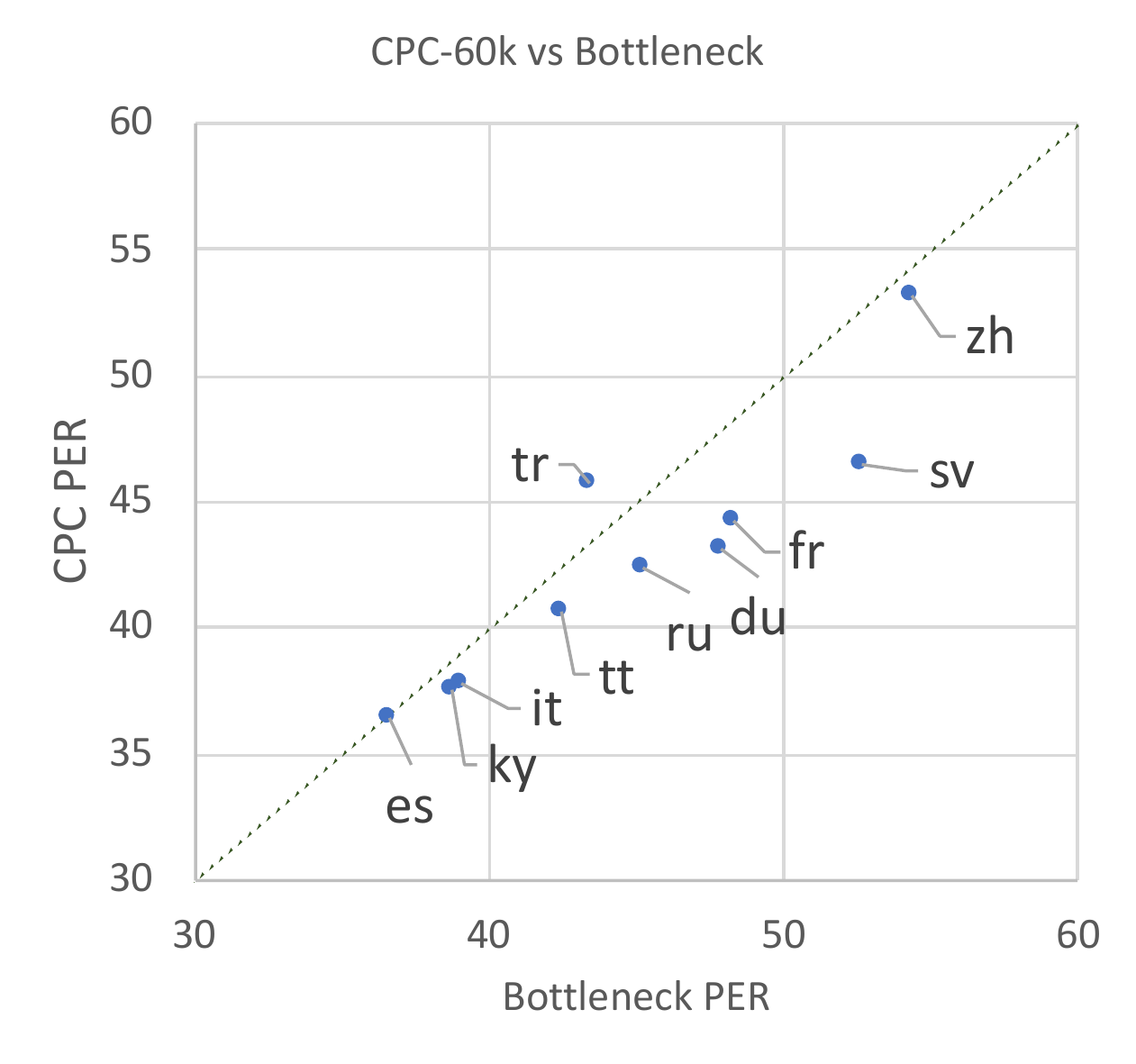}}
  \vspace{-1.3mm}
\caption{\textbf{CPC versus Bottleneck features.} The CPC features here have been trained on the 60Kh libri-light dataset.}
\label{fig:cpcvsbottle}
\end{figure}

Here, we present results on the CPC features trained on the recently released Libri-light 60K dataset\cite{kahn2020}. As seen in Table \ref{tab:cpc60k}, we now beat both the Bottleneck and Supervised features on all languages except one. The comparison between Bottleneck and CPC features is displayed in Figure \ref{fig:cpcvsbottle}.

\begin{table*}[h]
  \centering
  \begin{tabular}{lcccccccccccccc}
  \toprule
    Model & Pretraining & Frozen&  \texttt{du} & \texttt{es} & \texttt{fr}  & \texttt{it}  & \texttt{ky} & \texttt{ru} & \texttt{sv} & \texttt{tr} & \texttt{tt} & \texttt{zh}  && Avg \\
  \midrule
  Bottleneck~\cite{fer2017multilingually} & Babel-1070h 
         & Yes &    47.9&    36.6&    48.3&    39.0&    38.7&    45.2&    52.6&\bf 43.4& 42.5 & 54.3  && 44.9 \\
  Supervised & LS-100h                                  
         & Yes &\bf 42.4&\bf 36.4&    47.0&    40.5&    41.0&    43.6&    47.0&    48.5& 41.5 & 56.8 &&   44.5\\
  \midrule
    Modified CPC & LL-60K                               
         & Yes &    43.1&\bf 36.4&\bf 44.3&\bf 37.8&\bf 37.5&\bf 42.4&\bf 46.5&    45.7&\bf 40.6&\bf 53.2&&\bf 42.7 \\
  \bottomrule
  \end{tabular}
  \caption{
    {\bf Transfer of pre-trained phoneme features across languages.}
    Phone Error Rate on linear classification of phonemes based on pre-trained features on $60$kh of Libri-light, compared to multilingual bootleneck features (``Bottleneck'') trained on $1070$h from the Babel dataset and a supervised baseline trained on LibriSpeech 100h clean.
    The linear classifier is trained on the frozen features using $1$h of speech from the Common Voice database in different languages.
    We report Phone Error Rate.
    The languages are: 
    Dutch (\texttt{du}),
    Spanish (\texttt{es}),
    French (\texttt{fr}),
    Italian (\texttt{it}),
    Kyrgyz (\texttt{ky}),
    Russian (\texttt{ru}),
    Sweedish (\texttt{sv}),
    Turkish (\texttt{tr}),
    Tatar (\texttt{tt}) and
    Mandarin (\texttt{zh}).
  }
  \label{tab:cpc60k}
\end{table*}

\end{document}